\documentclass[10pt,twocolumn]{article}
\usepackage[margin=1in]{geometry}
\usepackage{amsmath}
\usepackage{abstract}
\usepackage{amsfonts,color}
\usepackage{amssymb}
\usepackage{graphicx}
\usepackage{hyperref}
\usepackage{authblk}
\begin{document}
\title{\bf{ Charged  wormholes   supported by 2-fluid immiscible matter particle}}

\author
{Safiqul Islam$^{1,\ast}$, Md Arif Shaikh$^{1,\dagger}$, Tapas Kumar Das$^{1,\ddagger}$ ,   Farook Rahaman$^{2,\mathsection}$ and Monsur Rahaman$^{2,**}$\\
$^1$Harish-Chandra Research Institute, HBNI, Chhatnag Road, Jhunsi, Allahabad-211019, India \\
$^2$Department of Mathematics, Jadavpur University, Kolkata 700032, India\\
\vspace{0.2cm}
$^\ast$safiqulislam@hri.res.in\\ $^\dagger$arifshaikh@hri.res.in\\ $^\ddagger $tapas@hri.res.in\\$^\mathsection$rahaman@associates.iucaa.in\\$^{**}$mansur.rahman90@gmail.com
}
\twocolumn[
\maketitle
\begin{onecolabstract}
We provide a 2-fluid immiscible matter source that supplies fuel to construct wormhole spacetime. The exact wormhole solutions are found in the model having, besides real matter or ordinary matter, some quintessence matter along with charge distribution. We intend to derive a general metric of a charged wormhole under some density profiles of galaxies that is also consistent with the observational profile of rotation curve of galaxies. We
have shown that the effective mass remains positive as well as the wormhole physics violate the null energy conditions. Some physical features are briefly discussed.
\end{onecolabstract}
]

\section{Introduction:}
~~~The existence of Lorentzian wormholes has been a speculative issue though there are views that it can exist at the planck scale of $10^{-35} $ metres \cite{j}. Though wormholes are designated as tunnels connecting two distinct universes or regions of the same universe but their actual physical significance was first proposed by Morris and Thorne \cite{a} and their possible locations in the galactic halo and central regions of galaxies further studied by Rahaman F. et al., \cite{b},\cite{l}. Exact solutions of the wormhole with extra fields such as scalar field and static electric charge have been studied by Kim S.W et al. \cite{g} and derived self-consistent solutions. We are acquainted with the fact that wormhole spacetimes are predictions of the GTR and need to be supported by observations. The modified extended metric for the charged wormhole model of Kim and Lee has been developed by Kuhfittig P.K.F. \cite{k}, to represent a charged wormhole that is compatible with quantum field theory. The field equations without the charge $Q$ have been studied by Buchdahl \cite{q} , who found physically meaningful solutions.

 {Kuhfittig P. K. F. et al. \cite{ak} studied the modified wormhole model in conjunction with a noncommutative-geometry background to overcome certain problems with traversability. Further in view of quantum field theory, the modified metric in \cite {k} is studied as a solution of the Einstein fields equations representing a charged wormhole. Stable phantom-energy wormholes admitting conformal motions is discusssed in \cite {am}. Here it is shown that the wormhole is stable to linearized radial perturbations whenever $−1.5 < \omega < −1$. The author has also studied macroscopic traversable wormholes with zero tidal forces inspired by noncommutative geometry \cite{ao} where it is shown that whenever the energy density describes a classical wormhole, the resulting solution is incompatible with quantum-field theory.

The Friedmann-Robertson-Walker model with a wormhole has been studied by Kim S. W. \cite{ab} and found the total matter to be nonexotic, while it is exotic in the static wormhole or inflating wormhole models, considering the matter distribution of the wormhole shape function as divided into two parts. The flare-out condition for the wormhole with the Einstein equation and the finiteness of the pressure is discussed in \cite{ac}. The possibility that inflation might provide a natural mechanism for the enlargement of Lorentzian wormholes of the Morris-Thorne type to macroscopic laws size has been explored by Roman T. A. \cite{ad}. An analytical electrically charged traversable wormhole solution for the Einstein–Maxwell-anti-dilaton theory as well as the deflection angle of a light-ray passing close to this wormhole has been discussed by Goulart P.\cite{ae}.
		
The traversability of asymptotically flat wormholes in Rastall gravity with phantom sources is studied in \cite{af}. It is also observed \cite{ag} that cosmic acceleration with traversable wormhole is possible without exotic matter like dark and phantom energy unless the scale factor of the universe obeys a power law dominated by a negative fractional parameter. 7-dimensional universe and the existence of a static traversable wormhole solutions in the form of the Lovelock gravity is studied by El-Nabulsi R. A.  \cite{ah}. The cosmological implications of a four-dimensional cosmology dominated by quintessence with a static traversable wormhole by means of an additional bimetric tensor in Einstein’s field
equations is discussed in \cite{an}. Extra-Dimensional cosmology with a traversable Wormhole has been studied in \cite{ap}. The author discuss many features of a higher-dimensional cosmology with a static traversable wormhole dominated by a variable effective cosmological constant and depending on the scale factor $a(t)$.

Hererra L. \cite{ai} has investigated the conditions under which general relativistic polytropes for anisotropic matter exhibit cracking and/or overturning. The dynamics of test particles in stable circular orbits around static and spherically symmetric wormholes in conformally symmetric spacetimes is discussed in \cite{aj}.

In this paper we thrive to derive a general metric of a charged wormhole under some density profiles of galaxies which should also be consistent with the observational profile of rotation curve of galaxies.

The stress in this paper is on the anisotropic matter distribution where we have considered a combined model of quintessence matter and ordinary matter along with charge distribution. A valid wormhole also exists under the isotropic condition where $p_t = p_r$. We iterate that such a combination could in fact support a wormhole in Einstein-Maxwell gravity. One could also like to invite the prospect of theoretical construction of such wormhole with the assumption of zero tidal forces.

  The aim of our paper is envisaged as follows:

  In section 2. we have solved the
 Einstein-Maxwell field equations pertaining to our proposed metric and deduce the wormhole solutions. Section 3. deals with the profile curve and the embedding diagram. The geodesics study in section 4. reveal the observational phenomena pertaining to gravity on the galactic scale which is attractive.  In section 5. and 6. we study the equilibrium conditions and the effective gravitational mass. Energy conditions and tidal forces are discussed in sections 7. The validity of our metric is further confirmed in our study of wormhole in the Galactic Halo region in section 8. The study ends with a concluding remark.

\section{A particular class of solutions:}
~~~We propose a general wormhole spacetime metric with charge for circular stable geodesic motion in the equatorial plane to be represented by the line element
\begin{eqnarray}
ds^{2}= -e^{\nu_{eff}}dt^{2}+(1-\frac{b_{eff}}{r})^{-1}dr^{2}\nonumber\\
+r^{2}(d\theta^{2}+sin^{2}\theta d\phi^{2}),
\end{eqnarray}

Here $\nu_{eff}$ and $b_{eff}$ are the effective redshift function and effective wormhole shape function having
functional dependence on the radial coordinate r and
\begin{eqnarray}
\nu_{eff}(r)=Log[(\frac{r}{r_s})^{l}+\frac{Q^2}{r^2}] \nonumber\\ e^{\lambda_{eff}(r)}=(1-\frac{b_{eff}}{r})^{-1}, \nonumber\\
 b_{eff}(r)= b(r)-\frac{Q^2}{r}
\end{eqnarray}
 The effective redshift function and the effective wormhole shape shape function are chosen in such a way that it does not violate the wormhole flare-out conditions as defined below. Also the spacetime is asymptotically flat, i.e $\frac{b_{eff}(r)}{r} \rightarrow 0$ as $|r| \rightarrow {\infty} $. The presence of charge inflicts a minor change in the graph of ${b_{eff}(r)}$ vs. r as against $b(r)$ vs. r, keeping in parity the desired wormhole flare-out conditions. We further show that for particular choice of the shape function and redshift function, the wormhole metric in the context does not violate the energy condition at or near the wormhole throat.

We consider the geometric units $G=c=1$, in the EM field equations and presumably set the wormhole spacetime with an effective and positive constant electric charge Q. Such a form of charge is valid in the wormhole metric as evident in \cite{g}. As the charge is static, there is no radiation by the fields. If $Q=0$, the metric reduces to the form observed by Rahaman F. et al. \cite{b} and if also $l=0$ alongwith $Q=0$, it reduces to the more generalized form of Morris-Thorne wormhole \cite{a}. Further if $b(r)=0$, it reduces to the RN black hole with a vanishing mass. Here $r_s$ denote the characteristic scale radius and $l$ is a constant defined by $l=2(v^{\phi})^{2}$, $v^{\phi}$ being the rotational velocity of a test particle in the same gravitational field in a stable circular orbit in the galactic halo region \cite{b}. The stability of such an orbit depends on the effective positive radial velocity and will be studied later.

The flare-out conditions need to be studied which prevent wormholes to be physical and making it open, thereby paving a way for traversibility. These are as :
(1) There should be no event horizon, hence the redshift function, $\nu_{eff}(r)$ should be finite.
(2) The wormhole shape function should obey the conditions, (i) $b_{eff}(r)\leqslant r$ for $r_{0}\leqslant r$, $r_{0}$ being the throat radius (ii) $b_{eff}(r_{0})=r_{0}$ (iii) $b_{eff}'(r_{0}) < 1$ (iv) $b_{eff}'(r) < \frac{b_{eff}(r)}{r}$ and (v) As $r \rightarrow r_0 $, $b_{eff}(r)$ approaches $2 M$, the Schwarzschildt mass \cite{j,r}, which is the mass function of the wormhole. Hence $b_{eff}(r)$ should be a positive function.

We propose that the matter sources consist of two non-interacting fluids, one being real matter in the form of perfect fluid and the second as anisotropic dark energy which is responsible for the acceleration of the universe \cite{f}.

The following self consistent Einstein-Maxwell equations for a charged fluid distribution is proposed as,
\begin{eqnarray}
G_{ab} = R_{ab} - \frac{1}{2} R g_{ab} =
8 \pi T^{eff}_{ab}\nonumber\\
= 8 \pi(T_{ab}+T^{de}_{ab}+T^{c}_{ab}),
\end{eqnarray}

where $T_{ab}$, $T^{de}_{ab}$ and $T^{c}_{ab}$ are the contributions to the effective energy-momentum tensor for matter fluid, exotic matter and charge respectively.

The most general energy-momentum tensor compatible with static
spherical symmetry for anisotropic distribution of matter ( the matter is exotic in nature and is a necessary ingredient for construction of wormhole, thereby violating the energy conditions ) is,
\begin{eqnarray}
T^{a(eff)}_{b} = (\rho^{eff} +p^{eff}_{t}) u^{a} u_{b} - p^{eff}_{t}g^a_{b} \nonumber\\+ (p^{eff}_{r}-p^{eff}_{t}) v^{a} v_{b},
\end{eqnarray}

where $\rho^{eff}$, $p^{eff}_{r}$, $p^{eff}_{t}$, $u_a$, and $v_{a}$ are, respectively,
effective matter-energy density,effective radial fluid pressure, effective transverse fluid pressure, four velocity, and radial four vector of
the fluid element. The case $p^{eff}_{t}=p^{eff}_{r}$, corresponds to the isotropic fluid when the anisotropic force vanishes.

In our consideration, the four velocity and radial four vector satisfy,
$u^{a} = e^{-\nu}{\delta^a_{0}}$, $u^{a} u_{a}=1$, $v^{a} = e^{-\lambda}{\delta^a_{1}}$, $v^{a} v_{a}=-1$.

From eqns. (3) and (4) we get the following set of eqns.,
\begin{equation}
T^{0}_{0}= \rho^{eff} = (\rho+\rho^{de}+\rho^{c})
\end{equation}

\begin{equation}
T^{1}_{1}= - p^{eff}_r = - (p+p^{de}_r+p^{c}_r)
\end{equation}

\begin{equation}
T^{2}_{2}= T^{3}_{3}= - p^{eff}_t = - (p+p^{de}_t+p^{c}_t),
\end{equation}

where $\rho^{c}$, $p^{c}_r$ and $p^{c}_t$ are the contributions due to the presence of charge.

We consider that the dark energy radial pressure is proportional to the dark energy density \cite{o},\cite{e} as,
\begin{equation}
p^{de}_r= -{\omega} \rho^{de}, ~~\frac{1}{3}<\omega<1,
\end{equation}

and the dark energy density is proportional to the mass density as,
\begin{equation}
\rho^{de} = n \rho, ~~~~n>0,
\end{equation}
 As the universe expands at a fixed rate, the curvature of spacetime is constant. Hence the density of dark energy which is responsible for the expansion of the universe, is also constant. Also from the first Friedmann equation, we find
$H^2 = (\frac{\dot{a}}{a})^2 = \frac{8 \pi G}{3} \rho$,
Here H is the Hubble parameter related with the expansion rate of the universe. This expansion rate is driven by mass density $\rho$.\\
We employ the following standard equation of state (EOS):
\begin{equation}
p = m \rho, ~~~~0 < m < 1,
\end{equation}
where m is a parameter corresponding to normal matter.

It needs to be verified if eqn.(1) satisfies the Einstein's equation self-consistently or nor.

The Einstein-Maxwell field equations with matter distribution as eqn (3), using eqns.(5)-(7), are analogous with the transformations,

\begin{equation}
e^{-\lambda_{eff}}(\frac{\lambda'_{eff}}{r}-\frac{1}{r^2})+\frac{1}{r^2} = 8 \pi (\rho+\rho^{de}+\rho^{c}),
\end{equation}

\begin{equation}
e^{-\lambda_{eff}}(\frac{\nu'_{eff}}{r}+\frac{1}{r^2})-\frac{1}{r^2} = 8 \pi (p+p^{de}_r+p^{c}_r),
\end{equation}

\begin{eqnarray}
\frac{1}{2}e^{-\lambda_{eff}}[\frac{1}{2}\nu_{eff}'^{2}+\nu_{eff}''-\frac{1}{2}\lambda_{eff}' \nu_{eff}'\nonumber\\ +\frac{1}{r}(\nu_{eff}'-\lambda_{eff}')]
= 8{\pi}(p+p^{de}_t+p^{c}_t),
\end{eqnarray}

Using eqns.(2),(8),(9) and (10), the above eqns.(11)-(13) reduce to,

\begin{equation}
~~~~~~~~~~~~~~~~\frac{b'(r)}{r^2}+ \frac{Q^2}{r^4}=8 \pi [(1+n) \rho +\rho^{c}],
\end{equation}

\begin{eqnarray}
\frac{Q^2}{r^4}-\frac{b(r)}{r^3}
-\frac{[2Q^2-lr^2(\frac{r}{r_s})^l][Q^2+r(r-b(r))]}{r^4[Q^2
+r^2(\frac{r}{r_s})^l]}\nonumber\\
=8 \pi [(1-n) \omega \rho+p^{c}_r],~~~~~~
\end{eqnarray}
and
\begin{eqnarray}
\frac{1}{4r^4 (Q^2+r^2 (\frac{r}{r_s})^l)^2}[4Q^6 +r^5 (\frac{r}{r_s})^{2l}\nonumber\\
(b(r)(2+l-l^2)+(l^2-b'(r)(l+2))r)\nonumber\\
+2Q^4 r(-2b(r)+r(2+2(\frac{r}{r_s})^l+l (\frac{r}{r_s})^l\nonumber\\+l^2 (\frac{r}{r_s})^l))
+Q^2 r^3 (\frac{r}{r_s})^l (-b(r)(2l^2+3l+6)\nonumber\\
+r(l^2 (2+(\frac{r}{r_s})^l)-2l ((\frac{r}{r_s})^l-2)\nonumber\\
-4(\frac{r}{r_s})^l-b'(r)(l+2)+8))]\nonumber\\=8 \pi [(m \rho+p^{de}_t)+p^{c}_t],
\end{eqnarray}

The matter terms due to charge are given by \cite{g},
\begin{equation}
\rho^{c} = p^{c}_r = p^{c}_t = \frac{Q^2}{8 \pi r^4 } ,
\end{equation}
Hence from eqns.(14) and (15) we find,
\begin{eqnarray}
b(r)= e^{\frac{(n+1)(-Log(r)+Log(Q^2+r^2 (\frac{r}{r_s})^l))}{\omega(n-1)}}\nonumber\\
+e^{\frac{(n+1)(-Log(r)+Log(Q^2+r^2 (\frac{r}{r_s})^l))}{\omega(n-1)}}\nonumber\\
\times \int_{1}^{r}[-e^{\frac{(n+1)(-Log(p)+Log(Q^2+p^2 (\frac{p}{r_s})^l))}{\omega(n-1)}}\nonumber\\
\times (n+1)(-2 Q^4 - 2Q^2 p^2 + m Q^2 p^2 (\frac{p}{r_s})^l\nonumber\\
+l p^4 (\frac{p}{r_s})^l)
\times (\omega (n-1)p^2 (Q^2+p^2(\frac{p}{r_s})^l))^{-1}]dp
\end{eqnarray}
where p is a variable. The effective wormhole shape function is found as,

\begin{eqnarray}
b_{eff}(r)= e^{\frac{(n+1)(-Log(r)+Log(Q^2+r^2 (\frac{r}{r_s})^l))}{\omega(n-1)}}-\frac{Q^2}{r}\nonumber\\
+e^{\frac{(n+1)(-Log(r)+Log(Q^2+r^2 (\frac{r}{r_s})^l))}{\omega(n-1)}}\nonumber\\
\times \int_{1}^{r}[-e^{\frac{(n+1)(-Log(p)+Log(Q^2+p^2 (\frac{p}{r_s})^l))}{\omega(n-1)}}\nonumber\\
\times (n+1)(-2 Q^4 - 2Q^2 p^2 + m Q^2 p^2 (\frac{p}{r_s})^l\nonumber\\
+l p^4 (\frac{p}{r_s})^l)
\times (\omega (n-1)p^2 (Q^2+p^2(\frac{p}{r_s})^l))^{-1}]dp\nonumber\\
\end{eqnarray}

From eqns. (14) and (17) we get expression for the energy density function as,

\begin{eqnarray}
\rho(r)=\frac{1}{8 \pi (1+n)r^2}\nonumber\\
 \times [\frac{(n+1)(2Q^4+2Q^2 r^2-l Q^2 r^2(\frac{r}{r_s})^l-lr^4 (\frac{r}{r_s})^l)}{\omega (n-1)r^2 (Q^2 + r^2 (\frac{r}{r_s})^l)}\nonumber\\
 +\frac{(n+1)}{(\omega(n-1))}e^{\frac{(n+1)(-Log(r)+Log(Q^2 + r^2 (\frac{r}{r_s})^l) )}{\omega(n-1)}}\nonumber\\
 \times(-\frac{1}{r}+\frac{\frac{lr^2 (\frac{r}{r_s})^{l-1}}{r_s}+2r (\frac{r}{r_s})^l}{Q^2+r^2 (\frac{r}{r_s})^l})\nonumber\\
 +\frac{(n+1)}{(\omega(n-1))}e^{\frac{(n+1)(-Log(r)+Log(Q^2 + r^2 (\frac{r}{r_s})^l) )}{\omega(n-1)}}\nonumber\\
  \times(-\frac{1}{r}+\frac{\frac{lr^2 (\frac{r}{r_s})^{l-1}}{r_s}+2r (\frac{r}{r_s})^l}{Q^2+r^2 (\frac{r}{r_s})^l})\nonumber\\
  \times \int_{1}^{r}(-e^{\frac{(n+1)(-Log(p)+Log(Q^2+p^2 (\frac{p}{r_s})^l))}{\omega(n-1)}}\nonumber\\
  \times (n+1)(-2 Q^4 - 2Q^2 p^2 + m Q^2 p^2 (\frac{p}{r_s})^l\nonumber\\
  +l p^4 (\frac{p}{r_s})^l)
  \times (\omega (n-1)p^2 (Q^2+p^2(\frac{p}{r_s})^l))^{-1})dp]
\end{eqnarray}

The following relations are evident,
\begin{eqnarray}
p(r) =\frac{m}{8 \pi (1+n)r^2}\nonumber\\
 \times [\frac{(n+1)(2Q^4+2Q^2 r^2-l Q^2 r^2(\frac{r}{r_s})^l-lr^4 (\frac{r}{r_s})^l)}{\omega (n-1)r^2 (Q^2 + r^2 (\frac{r}{r_s})^l)}\nonumber\\
  +\frac{(n+1)}{(\omega(n-1))}e^{\frac{(n+1)(-Log(r)+Log(Q^2 + r^2 (\frac{r}{r_s})^l) )}{\omega(n-1)}}\nonumber\\
  \times(-\frac{1}{r}+\frac{\frac{lr^2 (\frac{r}{r_s})^{l-1}}{r_s}+2r (\frac{r}{r_s})^l}{Q^2+r^2 (\frac{r}{r_s})^l})\nonumber\\
  +\frac{(n+1)}{(\omega(n-1))}e^{\frac{(n+1)(-Log(r)+Log(Q^2 + r^2 (\frac{r}{r_s})^l) )}{\omega(n-1)}}\nonumber\\
   \times(-\frac{1}{r}+\frac{\frac{lr^2 (\frac{r}{r_s})^{l-1}}{r_s}+2r (\frac{r}{r_s})^l}{Q^2+r^2 (\frac{r}{r_s})^l})\nonumber\\
   \times \int_{1}^{r}(-e^{\frac{(n+1)(-Log(p)+Log(Q^2+p^2 (\frac{p}{r_s})^l))}{\omega(n-1)}}\nonumber\\
   \times (n+1)(-2 Q^4 - 2Q^2 p^2 + l Q^2 p^2 (\frac{p}{r_s})^l\nonumber\\
   +l p^4 (\frac{p}{r_s})^l)
   \times (\omega (n-1)p^2 (Q^2+p^2(\frac{p}{r_s})^l))^{-1})dp]
\end{eqnarray}

\begin{eqnarray}
\rho^{de} =\frac{n}{8 \pi (1+n)r^2}\nonumber\\
 \times [\frac{(n+1)(2Q^4+2Q^2 r^2-l Q^2 r^2(\frac{r}{r_s})^l-lr^4 (\frac{r}{r_s})^l)}{\omega (n-1)r^2 (Q^2 + r^2 (\frac{r}{r_s})^l)}\nonumber\\
  +\frac{(n+1)}{(\omega(n-1))}e^{\frac{(n+1)(-Log(r)+Log(Q^2 + r^2 (\frac{r}{r_s})^l) )}{\omega(n-1)}}\nonumber\\
  \times(-\frac{1}{r}+\frac{\frac{lr^2 (\frac{r}{r_s})^{l-1}}{r_s}+2r (\frac{r}{r_s})^l}{Q^2+r^2 (\frac{r}{r_s})^l})\nonumber\\
  +\frac{(n+1)}{(\omega(n-1))}e^{\frac{(n+1)(-Log(r)+Log(Q^2 + r^2 (\frac{r}{r_s})^l) )}{\omega(n-1)}}\nonumber\\
   \times(-\frac{1}{r}+\frac{\frac{lr^2 (\frac{r}{r_s})^{l-1}}{r_s}+2r (\frac{r}{r_s})^l}{Q^2+r^2 (\frac{r}{r_s})^l})\nonumber\\
   \times \int_{1}^{r}(-e^{\frac{(n+1)(-Log(p)+Log(Q^2+p^2 (\frac{p}{r_s})^l))}{\omega(n-1)}}\nonumber\\
   \times (n+1)(-2 Q^4 - 2Q^2 p^2 + l Q^2 p^2 (\frac{p}{r_s})^l\nonumber\\
   +l p^4 (\frac{p}{r_s})^l)
   \times (\omega (n-1)p^2 (Q^2+p^2(\frac{p}{r_s})^l))^{-1})dp]
\end{eqnarray}

\begin{eqnarray}
 p^{de}_r =\frac{-n \omega}{8 \pi (1+n)r^2}\nonumber\\
 \times [\frac{(n+1)(2Q^4+2Q^2 r^2-l Q^2 r^2(\frac{r}{r_s})^l-lr^4 (\frac{r}{r_s})^l)}{\omega (n-1)r^2 (Q^2 + r^2 (\frac{r}{r_s})^l)}\nonumber\\
  +\frac{(n+1)}{(\omega(n-1))}e^{\frac{(n+1)(-Log(r)+Log(Q^2 + r^2 (\frac{r}{r_s})^l) )}{\omega(n-1)}}\nonumber\\
  \times(-\frac{1}{r}+\frac{\frac{lr^2 (\frac{r}{r_s})^{l-1}}{r_s}+2r (\frac{r}{r_s})^l}{Q^2+r^2 (\frac{r}{r_s})^l})\nonumber\\
  +\frac{(n+1)}{(\omega(n-1))}e^{\frac{(n+1)(-Log(r)+Log(Q^2 + r^2 (\frac{r}{r_s})^l) )}{\omega(n-1)}}\nonumber\\
   \times(-\frac{1}{r}+\frac{\frac{lr^2 (\frac{r}{r_s})^{l-1}}{r_s}+2r (\frac{r}{r_s})^l}{Q^2+r^2 (\frac{r}{r_s})^l})\nonumber\\
   \times \int_{1}^{r}(-e^{\frac{(n+1)(-Log(p)+Log(Q^2+p^2 (\frac{p}{r_s})^l))}{\omega(n-1)}}\nonumber\\
   \times (n+1)(-2 Q^4 - 2Q^2 p^2 + l Q^2 p^2 (\frac{p}{r_s})^l\nonumber\\
   +l p^4 (\frac{p}{r_s})^l)
   \times (\omega (n-1)p^2 (Q^2+p^2(\frac{p}{r_s})^l))^{-1})dp]
\end{eqnarray}
and a big expression for the dark energy tangential pressure as,

\begin{eqnarray}
 p^{de}_t =\frac{1}{32 \pi r^4 (Q^2+r^2 (\frac{r}{r_s})^l)^2}[4Q^6 +r^5 (\frac{r}{r_s})^{2l}\nonumber\\
 (b(r)(2+l-l^2)+(l^2-b'(r)(l+2))r)\nonumber\\
 +2Q^4 r(-2b(r)+r(2+2(\frac{r}{r_s})^l+l (\frac{r}{r_s})^l\nonumber\\+l^2 (\frac{r}{r_s})^l))
 +Q^2 r^3 (\frac{r}{r_s})^l (-b(r)(2l^2+3l+6)\nonumber\\
 +r(l^2 (2+(\frac{r}{r_s})^l)-2l ((\frac{r}{r_s})^l-2)\nonumber\\
 -4(\frac{r}{r_s})^l-b'(r)(l+2)+8))]-\frac{m}{8 \pi (1+n)r^2}\nonumber\\
  \times [\frac{(n+1)(2Q^4+2Q^2 r^2-l Q^2 r^2(\frac{r}{r_s})^l-lr^4 (\frac{r}{r_s})^l)}{\omega (n-1)r^2 (Q^2 + r^2 (\frac{r}{r_s})^l)}\nonumber\\
   +\frac{(n+1)}{(\omega(n-1))}e^{\frac{(n+1)(-Log(r)+Log(Q^2 + r^2 (\frac{r}{r_s})^l) )}{\omega(n-1)}}\nonumber\\
   \times(-\frac{1}{r}+\frac{\frac{lr^2 (\frac{r}{r_s})^{l-1}}{r_s}+2r (\frac{r}{r_s})^l}{Q^2+r^2 (\frac{r}{r_s})^l})\nonumber\\
   +\frac{(n+1)}{(\omega(n-1))}e^{\frac{(n+1)(-Log(r)+Log(Q^2 + r^2 (\frac{r}{r_s})^l) )}{\omega(n-1)}}\nonumber\\
    \times(-\frac{1}{r}+\frac{\frac{lr^2 (\frac{r}{r_s})^{l-1}}{r_s}+2r (\frac{r}{r_s})^l}{Q^2+r^2 (\frac{r}{r_s})^l})\nonumber\\
    \times \int_{1}^{r}(-e^{\frac{(n+1)(-Log(p)+Log(Q^2+p^2 (\frac{p}{r_s})^l))}{\omega(n-1)}}\nonumber\\
    \times (n+1)(-2 Q^4 - 2Q^2 p^2 + l Q^2 p^2 (\frac{p}{r_s})^l\nonumber\\
    +l p^4 (\frac{p}{r_s})^l)
    \times (\omega (n-1)p^2 \nonumber\\(Q^2+p^2(\frac{p}{r_s})^l))^{-1})dp]
    -\frac{Q^2}{8 \pi r^4 }
\end{eqnarray}

\subsection {$n=\frac{1}{2}$:}
The wormhole shape function is plotted below [Assuming $Q^2=0.1, \omega=\frac{2}{3}, l=0.05, r_{s}=1$].
\begin{figure}[htbp]
\centering
\includegraphics[scale=.8]{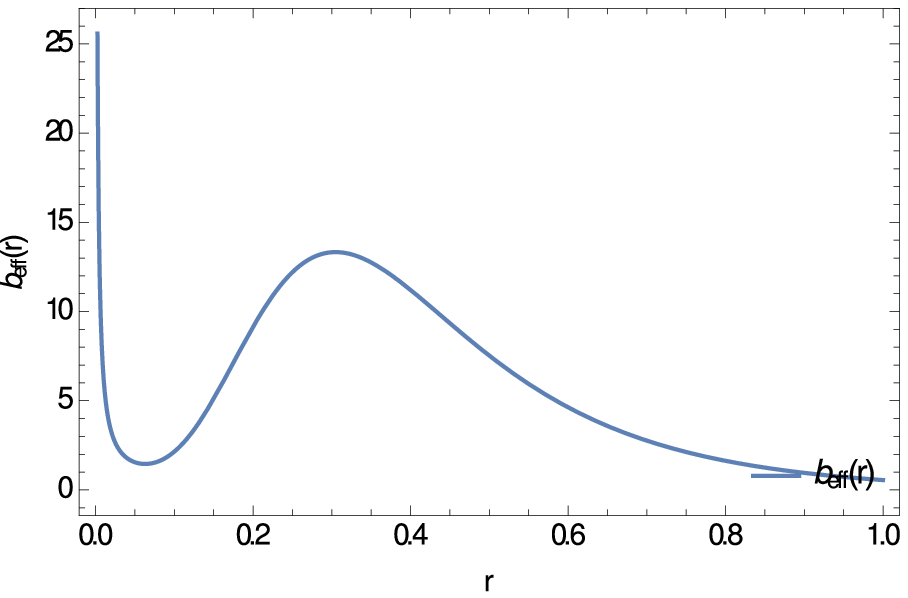}
\caption{The shape function of the wormhole against $r(k.p.c)$.}
\end{figure}

\begin{figure}[htbp]
\centering
\includegraphics[scale=.8]{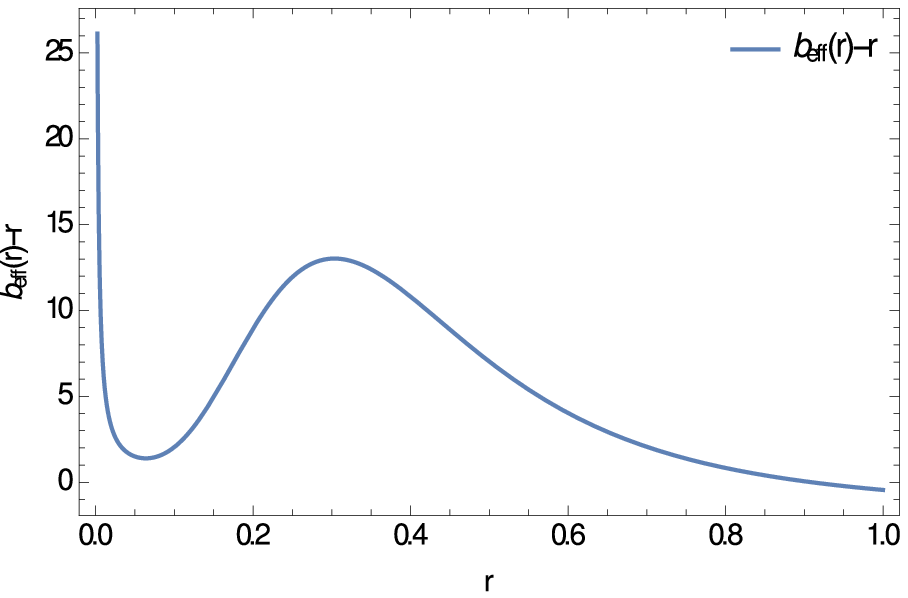}
\caption{The throat occurs where $b_{eff}(r)-r$ cuts the r-axis.}
\end{figure}

\begin{figure}[htbp]
\centering
\includegraphics[scale=.8]{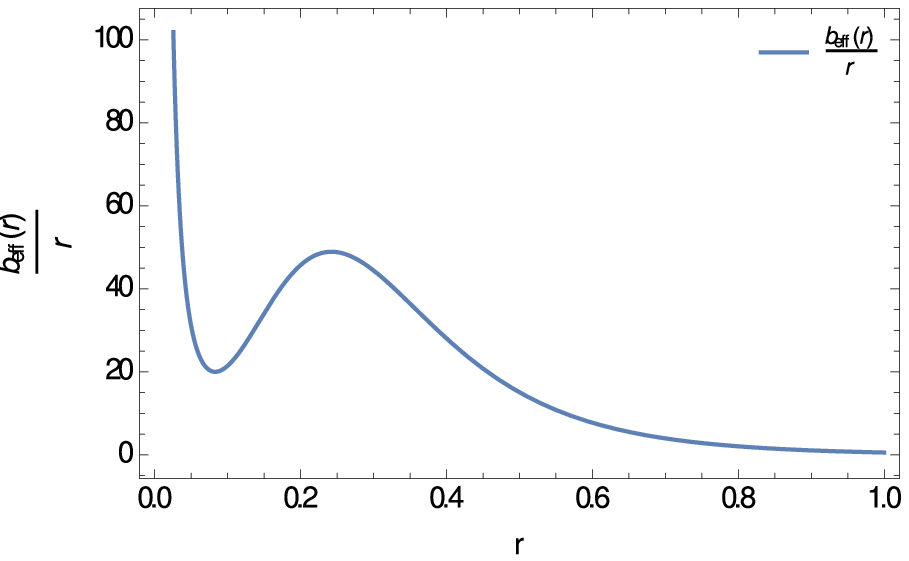}
\caption{$\frac{b_{eff}(r)}{r}$ is plotted against $r(k.p.c)$.}
\end{figure}

\begin{figure}[htbp]
\centering
\includegraphics[scale=.8]{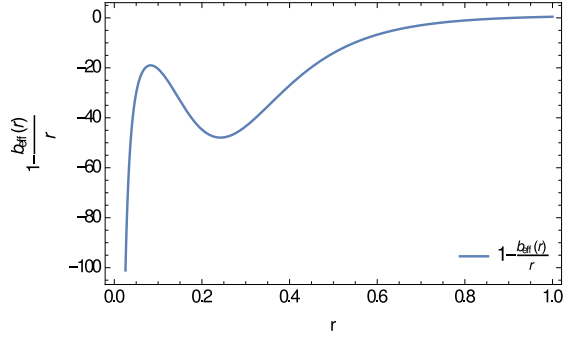}
\caption{$(1-\frac{b_{eff}(r)}{r})$ is plotted against $r(k.p.c)$.}
\end{figure}

It is evident from the fig.1, that the throat occurs at $r_{0}=0.9092$. The flare-out conditions, $b_{eff}(0.9092)=0.9092$ and $b_{eff}'(0.9092)=-4.92297<1$ are also satisfied. As the wormhole shape function $b_{eff}(r)$ and the energy momentum tensor $T^{a}_b$ self-consistently satisfy the E-M equations, hence the wormhole spacetime metric in eqn.(1) is valid.


The electric field \cite{g} is given below and plotted graphically in fig(5),
\begin{eqnarray}
E(r)=\frac{Q}{r^2} \times \sqrt{g_{tt} g_{rr}}\nonumber\\
=\frac{Q}{r^2} \times \sqrt{ \frac{(\frac{r}{r_s})^{l}+\frac{Q^2}{r^2}}{(1-\frac{b(r)}{r}+\frac{Q^2}{r^2})}},
\end{eqnarray}

\begin{figure}[htbp]
\centering
\includegraphics[scale=.8]{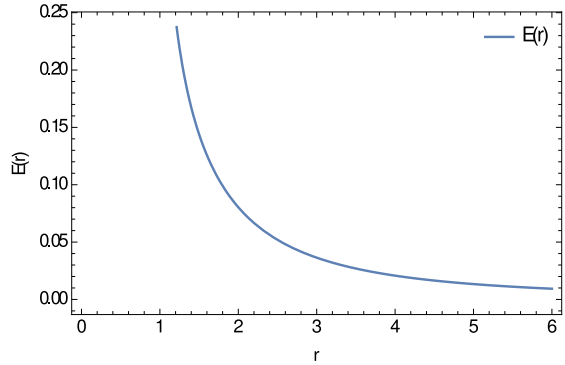}
\caption{The elctric field is plotted against $r(k.p.c)$.}
\end{figure}

The wormhole metric is valid throughout the galactic halo region. Also the test particle and the wormhole itself are in the same gravitational field of the galactic halo. Hence, the flat rotation curve for the circular stable geodesic motion of a test particle on the outer regions of the galactic halo and in the equatorial plane yields the tangential velocity \cite{h},\cite{i},\cite{n} as,
\begin{equation}
(v^{\phi})^2 = \frac{r}{2} \nu'(r),
\end{equation}

Using eqn.(2) we find,
\begin{equation}
(v^{\phi})^2 = \frac{l (\frac{r}{r_s})^l - \frac{2 Q^2}{r^2}}{2[(\frac{r}{r_s})^l +\frac{Q^2}{r^2}]},
\end{equation}

The velocity profile is shown in fig.(6) which is also consistent as per observations,
\begin{figure}[htbp]
\centering
\includegraphics[scale=0.8]{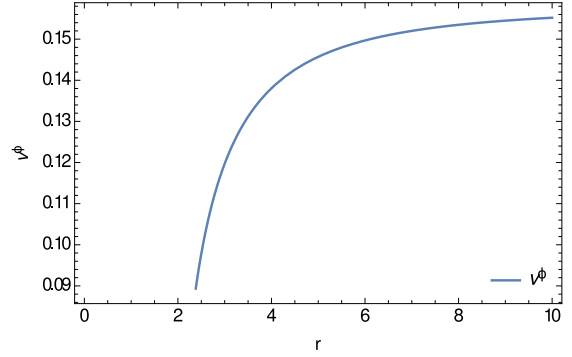}
\caption{The velocity of galactic rotation curve is shown against $r(k.p.c)$.}
\end{figure}

However, if $l=0$, then $Q$ should also vanish simultaneously for a static and uncharged wormhole solution. For a positive circular rotational velocity, when $l$ is considered as a parameter dependent on $r$ we find,
\begin{equation}
l \geq \frac{Lambert W[\frac{2 Q^2 Log (\frac{r}{r_s})}{r^2}]}{Log (\frac{r}{r_s})} ,
\end{equation}
where the Lambert W-function also called omega function is the inverse function of $f(W)= W e^W$.

For static and charged wormhole, l(r) decreases as $r$ increases and is shown below,
\begin{figure}[htbp]
\centering
\includegraphics[scale=.8]{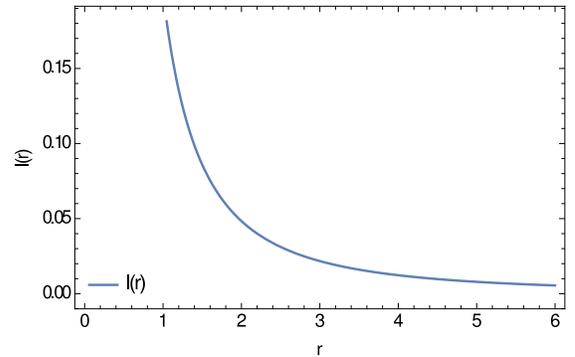}
\caption{l(r) is shown against $r$.}
\end{figure}

\section{Profile curve and embedding diagram of the wormhole: }
The profile curve is of the wormhole is defined by \cite{m}, \cite{a},
\begin{eqnarray}
\frac{dz}{dr}= \pm \frac{1}{\sqrt{r/b_{eff}(r)-1}}\nonumber\\
= \pm [r \times (e^{\frac{(n+1)(-Log(r)+Log(Q^2+r^2 (\frac{r}{r_s})^l))}{\omega(n-1)}}-\frac{Q^2}{r}\nonumber\\
+e^{\frac{(n+1)(-Log(r)+Log(Q^2+r^2 (\frac{r}{r_s})^l))}{\omega(n-1)}}\nonumber\\
\times \int_{1}^{r}[-e^{\frac{(n+1)(-Log(p)+Log(Q^2+p^2 (\frac{p}{r_s})^l))}{\omega(n-1)}}\nonumber\\
\times (n+1)(-2 Q^4 - 2Q^2 p^2 + m Q^2 p^2 (\frac{p}{r_s})^l\nonumber\\
+l p^4 (\frac{p}{r_s})^l)
\times (\omega (n-1)p^2 (Q^2\nonumber\\+p^2(\frac{p}{r_s})^l))^{-1}]dp)^{-1}-1 ]^{-\frac{1}{2}},
\end{eqnarray}

The embedding diagram for a particular choice of parameters $Q^2=0.1, n=\frac{1}{2},l=0.05,\omega=\frac{2}{3},r_s=1$, is obtained by rotating the profile curve about the z-axis as follows,

\begin{figure}[htbp]
\centering
\includegraphics[scale=.6]{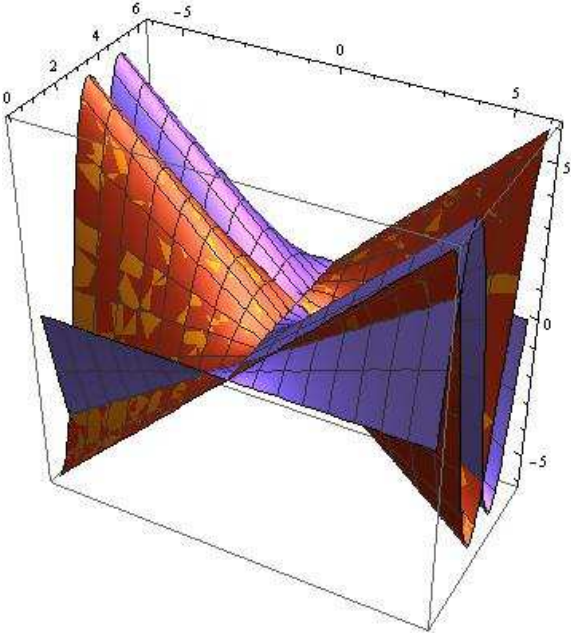}
\caption{Embedding diagram of the wormhole.}
\end{figure}

We observe that the wormhole is valid in a particular region upto $r=2 ~k.p.c$, the throat occuring at $r=0.9092~k.p.c$ from it's center.

According to Morris and Thorne \cite{a}, the r-coordinate
is ill-behaved near the throat, but proper radial distance must be well behaved everywhere i.e. we must require that
l(r) is finite throughout the spacetime.
The proper radial distance L(r) from the throat to a point
outside is
\begin{eqnarray}
L(r)= \pm \int_{r^{+}_{0}}^r \frac{dr}{\sqrt{1-\frac{b_{eff}(r)}{r}}}
\end{eqnarray}
which is always positive taking $r_0=0.9092$.

\section{Geodesic equations:}

We now study the geodesic equation for a test particle placed at some radius $r_0$. The radial equation is found as,
\begin{eqnarray}
\frac{d^2 r}{ds^2} = - [\frac{4 Q^2 + r(3 b(r)-(b'(r)+2)r)}{2r}(\frac{d\theta}{ds})^2 \nonumber\\
+ \frac{c^2_{2}(Q^2+r(r-b(r)))(-2Q^2+lr^2(\frac{r}{r_s})^l)}{2r (Q^2+r^2 (\frac {r}{r_s})^l)} \nonumber\\
+ \frac{c^2_{1} (Q^2+r(b(r)-r) cosec^2{\theta})}{r^5}\nonumber\\
+\frac{(1-b(r)+\frac{Q^2}{r^2})(2Q^2+r (r b'(r)-b(r)))}{2r(Q^2+r(r-b(r)))}\nonumber\\
\times (1+\frac{c^2_{2}}{\frac{Q^2}{r^2}+(\frac{r}{r_s})^l}-\frac{c^2_{1} cosec^2{\theta}}{r^2})]_{r=r_{0}},
\end{eqnarray}

where $c_1$ and $c_2$ are arbitrary constants. From the equations we find that $\frac{d^2 r}{ds^2}=-0.157673$, is negative at $r_0$ [considering $\theta=\frac{\pi}{2}$ and $\frac{d \theta}{ds}=1$].
Also the quantity in the square bracket is positive.
Hence the centrepetal force is directed towards the center of rotation indicating that the motion is stable.
Thus particles are attracted towards the center. This result confirms with the observational
phenomena pertaining to gravity on the galactic scale which is attractive (clustering, structure formation etc.) \cite{n}.

\section{Equilibrium conditions: }

Following \cite{p}, we write the TOV Eq. for an anisotropic fluid distribution, in the following form
\begin{eqnarray}
-\frac{M_G (\rho^{eff}+p_r^{eff}}{r^2}e^{\frac{\lambda-\nu}{2}}-\frac{dp_r^{eff}}{dr}\nonumber\\
+\frac{2}{r} (p_t^{eff}-p_r^{eff} )
+\sigma E(r)=0,
\end{eqnarray}

where $M_G=M_G(r)$ is the effective gravitational mass within the
radius $r$ and is given by
\begin{equation}
M_G(r)=\frac{1}{2}r^2e^{\frac{\nu-\lambda}{2}}\nu^{\prime},
\end{equation}
which can easily be derived from the Tolman-Whittaker formula and
the Einstein's field equations. Obviously, the modified TOV
equation (32) describes the equilibrium condition for the wormhole
subject to gravitational ($F_g$) and hydrostatic ($F_h$) plus
another force due to the anisotropic nature ($F_a$) of the matter
comprising the wormhole. Therefore, for equilibrium the above Eq. can be written as
\begin{equation}
 F_g+ F_h + F_a + F_e=0,
\end{equation}
where,
\begin{equation}
F_g =-\frac{\nu^\prime}{2}(\rho^{eff} +p_r^{eff}),
\end{equation}

\begin{equation}
F_h = -\frac{dp_r^{eff}}{dr}
\end{equation}

\begin{equation}
F_a=\frac{2}{r}(p_t^{eff} -p_r^{eff})
\end{equation}

\begin{equation}
F_e = \sigma(r) E(r)
\end{equation}

where the proper charge density $\sigma(r)$ is given by,
\begin{equation}
(r^2 E(r))'= 4 \pi r^2 \sigma(r) e^{\frac{\lambda(r)}{2}}
\end{equation}

The profiles of $F_g$, $F_h$, $F_e$ and $F_a$ for our chosen source are
shown in Fig 9. The figure indicates that equilibrium stage can
be achieved due to the combined effect of pressure anisotropic, electrical, gravitational and hydrostatic forces. It is to be distinctly noted that the anisotropic force is balanced by the combined effects of the forces electrical, gravitational and hydrostatic forces.

\begin{figure}[htbp]
\centering
\includegraphics[scale=.8]{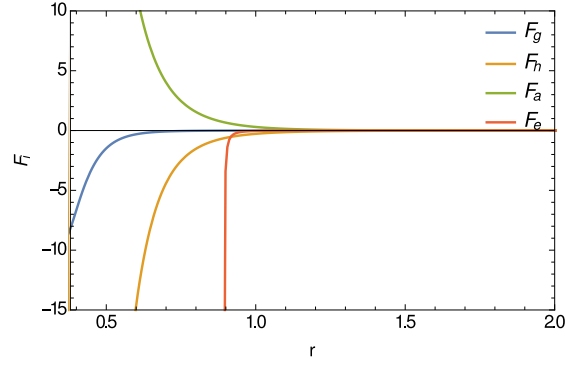}
\caption{The various forces acting on the system in equilibrium are show against $r(k.p.c)$.}
\end{figure}

\section{Effective gravitational mass}
In our model the effective gravitational mass, in terms of the
effective energy density $\rho^{eff}$, can be expressed as

\begin{eqnarray}
M^{eff}&=& 4\pi\int^{R}_{r_0}(\rho^{eff}) dr \nonumber\\
       &=& 4\pi\int^{R}_{r_0}(\rho+\rho^{de}+\rho^c) r^2 dr
\end{eqnarray}

The effective mass of the wormhole of throat
radius, say, $r_0 = 0.9092~k.p.c$ upto $10~k.p.c$ is obtained as $M^{eff}=0.637424~k.p.c$.
One can observe that the active gravitational mass $M^{eff}$ of the
wormhole is positive. This indicates that seen from the
Earth, it is not possible to alienate the gravitational
nature of a wormhole from that of a compact mass in the
galaxy.

\section{Energy Conditions and tidal forces: }
We observe that $\rho+p<0$, which is in violation of the null energy condition to support wormholes.

\begin{figure}[htbp]
\centering
\includegraphics[scale=.8]{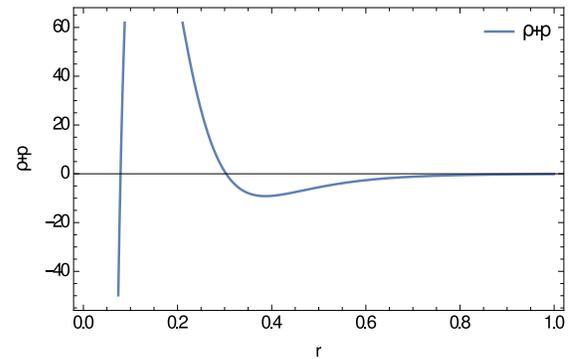}
\caption{Null energy condition plotted against against $r$ in k.p.c.}
\end{figure}

We observe that the wormhole has two distinct regions, in the region $0.3<r<2~~k.p.c$, $\rho+p<0$,
whereas in the region $2<r<6~~k.p.c$, $\rho+p \rightarrow 0$ as $r \rightarrow \infty$

\section{Density profile in the Galactic Halo region:}

We further consider the density profile of galaxies and their clusters in this region which takes the following form \cite{c,d},
\begin{equation}
\rho(r) = \frac{\rho_s}{\frac{r}{r_s}(1+\frac{r}{r_s})^{2}},
\end{equation}
where $r_{s}$ as defined earlier and $\rho_s$ is the corresponding density, and search for further validity of our wormhole metric.

Hence the wormhole shape function is obtained from eqn.(14) as,

\begin{eqnarray}
b_{eff}(r)= 8 \pi (1+n) \rho_{s} r^{3}_{s}\times [ln(1+\frac{r}{r_s})\nonumber\\
+\frac{1}{1+\frac{r}{r_s}}]
-\frac{Q^2}{r},
\end{eqnarray}
when,
\begin{equation}
b(r)=8 \pi (1+n) \rho_{s} r^{3}_{s}\times [ln(1+\frac{r}{r_s})+\frac{1}{1+\frac{r}{r_s}}],
\end{equation}

We now consider the charge to be dependent on the wormhole shape function as \cite{k},
\begin{equation}
\frac{2 Q^2}{r^2} = b_{eff}'(r),
\end{equation}
that is when,
\begin{equation}
\frac{Q^2}{r^2} = b'(r),
\end{equation}
Thus we get from eqn.(14),
\begin{equation}
\frac{2 Q^2}{r^4} = 8 \pi [\frac{(1+n) \rho_s}{\frac{r}{r_s}(1+\frac{r}{r_s})^{2}}+\rho^c],
\end{equation}

Eqn.(17) and (46) implies,
\begin{equation}
\frac{Q^2}{r^4} = 8 \pi [\frac{(1+n) \rho_s}{\frac{r}{r_s}(1+\frac{r}{r_s})^{2}}],
\end{equation}
and hence the wormhole charge within the throat radius is given by,

\begin{equation}
Q^2=8 \pi r_{0}^4    [\frac{(1+n)\rho_s}{\frac{r_{0}}{r_s}(1+\frac{r_{0}}{r_s})^{2}}
],
\end{equation}

which is constant for a suitable choice of positive parameter, n.
The electric field is given by \cite{g}
\begin{eqnarray}
E(r)=\frac{Q}{r^2}\sqrt{g_{tt} g_{rr}}\nonumber\\
=\sqrt{8 \pi r^4    [\frac{(1+n)\rho_s}{\frac{r}{r_s}(1+\frac{r}{r_s})^{2}}
]
\times \frac{(\frac{r}{r_s})^{l}+\frac{Q^2}{r^2}}{(1-\frac{b(r)}{r}+\frac{Q^2}{r^2})}},
\end{eqnarray}

The wormhole shape function for suitable choice of parameters $Q, \rho_{s}, n$ and $r_{s}$ is plotted below,
\begin{figure}[htbp]
\centering
\includegraphics[scale=.8]{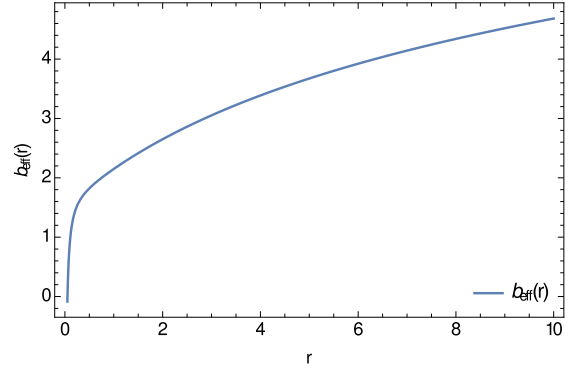}
\caption{The shape function of the wormhole against $r$($ k.p.c$).}
\end{figure}

The other conditions are plotted in fig.(12) and fig.(13),
\begin{figure}[htbp]
\centering
\includegraphics[scale=.8]{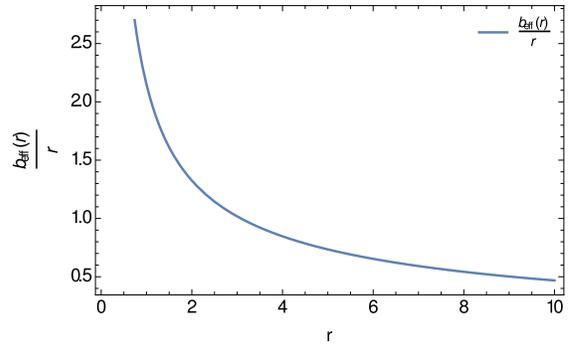}
\caption{The asymptotic behaviour of shape function against $r(k.p.c)$.}
\end{figure}

\begin{figure}[htbp]
\centering
\includegraphics[scale=.8]{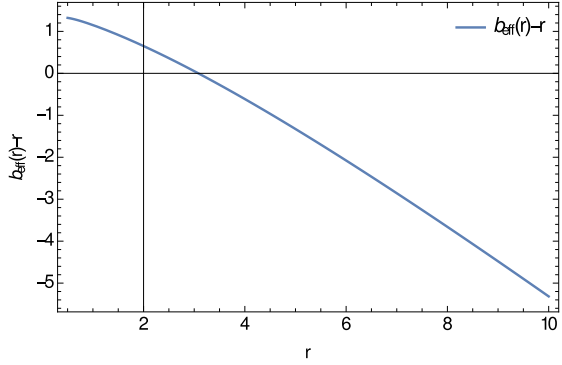}
\caption{The throat occurs where $b_{eff}(r)-r$ cuts the r-axis.}
\end{figure}

Fig.(13) indicates that the throat of the wormhole occurs at $r=3.0819$. Also as $b_{eff}'(3.0819)=0.359322<1$, the conditions for a valid wormhole are satisfied.




\section{Final Remarks}
~ The possibility of generalizing the wormholes discussed here to wormholes with (slow) rotation has been discussed in \cite {da}. In the case of slow rotation a small nondiagonal polar-angle-time $(t \phi)$ component of metric is included. The $(t \phi)$ Einstein equation can be written using the tetrad components introduced with the basic 1-forms ${\omega}^a = e^a_{\mu} dx^{\mu}$ \cite {h} and using $\phi = \frac{\nu_{eff}}{2}$ as,
\begin{eqnarray}
\omega^0 = e^{\varPhi} dt, \omega^1 = (d \phi + h dt) r sin \theta,\nonumber\\ \omega^2 = d \rho, \omega^3 = r d\theta,
\end{eqnarray}
Hence eqn.(1) is reduced to the self-sustained static
wormhole metric,
\begin{eqnarray}
ds^{2}= -e^{2 \varPhi} dt^{2}+d \rho^{2}
+r^{2}[d\theta^{2}+sin^{2}\theta (d\phi^{2}\nonumber\\+2 h d\phi dt)] ,
\end{eqnarray}	
where $h(\rho, \theta)$ is the angular velocity of rotation. The static metric $\varPhi(\rho)$, $r(\rho)$ is assumed to exist as a self-consistent solution of static semiclassical Einstein equations and $h$ is considered to be arbitrarily small.

We can also extend the original model of wormhole to a more generalized cosmological traversable wormholes by introducing the fifth coordinate $z$ as detailed in \cite{db}, such that
\begin{equation}
\frac{dz}{dr} = \pm [\frac{r}{b(r)}-1]^{-\frac{1}{2}}.
\end{equation}
Thus using $\phi = \frac{\nu_{eff}}{2}$ eqn.(1) reduces to the 5-dimensional metric as,
\begin{eqnarray}
ds^2 = -e^{2 \phi(r)} dt^2 + dz^2 + dr^2 \nonumber\\+ r^2 (d\theta^2 + sin^2 \theta d\phi^2 ).
\end{eqnarray}

It is known that the presence of a nonminimal interaction between dark matter and dark energy may lead to
a violation of the null energy condition and to the formation of a configuration with nontrivial topology like a wormhole \cite{dc}. Here the violation of N.E.C takes place only in the inner high-density regions of the configuration both between $0.3<r<2~~k.p.c$ and $2<r<6~~k.p.c$.

The present observational data also suggest an exotic form of dark energy with equation of state $p < -\rho$, violating the weak energy condition \cite{dd},\cite{de}. This violation allows for exotic solutions of general relativity such as wormholes and warp drives and the possibility of time travel associated with them. Dark energy with $\omega = \frac{p}{\rho} < -1$, is called super-quintessence or phantom energy \cite{df}.

We observe via eqn.(1) that as the metric components are positive, the spacetime metric is devoid of any horizon resulting in a positive wormhole mass. Also the shape function is dependent on the charge $Q$, the characteristic scale length $r_s$, the corresponding density $\rho_s$ and the positive constant $n$. Moreover for traversable wormhole the charge is considered sufficiently small yet positive for feasible geometric configuration. The validity of the metric is further confirmed by our extended study of wormholes in the galactic halo region. But whatever be the value of the charge the wormhole throat in our model occurs at a distance $0.9092~k.p.c$ from the center whereas in the Galactic Halo region the throat occurs at a distance $3.0819~k.p.c$ from the center for suitable choices of the above values. Hence we may conclude that any charge inside the wormhole is static in nature and could not effect any change in the wormhole configurations.

We iterate that such a metric is useful for wormhole study in other regions of the galaxy under different forms of gravity as well. Such study is in progress.

\section{Acknowledgments}
~~~FR would like to thank the authorities of the Inter-
University Centre for Astronomy and Astrophysics,
Pune, India for providing the research facilities. FR and
MR are also thankful to DST-SERB and UGC for financial support. We are grateful to the referee for his valuable comments.

\end{document}